\documentclass[letterpaper,11pt]{article}
\usepackage{jheppub}
\usepackage{physics}
\usepackage{relsize}
\usepackage{tikz}
\usepackage{pgfplots}
\usepackage{comment}

\usepackage[utf8]{inputenc}

\title{Conformal Boundary Conditions from Cutoff AdS$_{3}$}
\author[a]{Evan Coleman,}
\author[a,b]{Vasudev Shyam}
\affiliation[a]{Stanford Institute for Theoretical Physics and Department of Physics, Stanford University, Stanford, CA 94305, USA}
\affiliation[b]{Perimeter Institute for Theoretical Physics, 31 Caroline St. N, N2L 2Y5, Waterloo ON, Canada}
\date{September 2020}

\begin{document}

\abstract{
	We construct a particular flow in the space of 2D Euclidean QFTs on a torus, which we argue is dual to a class of solutions in 3D Euclidean gravity with conformal boundary conditions. This new flow comes from a Legendre transform of the kernel which implements the $T\bar{T}$ deformation, and is motivated by the need for boundary conditions in Euclidean gravity to be elliptic, i.e. that they have well-defined propagators for metric fluctuations. We demonstrate equivalence between our flow equation and variants of the Wheeler de-Witt equation for a torus universe in the so-called Constant Mean Curvature (CMC) slicing. We derive a kernel for the flow, and we compute the corresponding ground state energy in the low-temperature limit. Once deformation parameters are fixed, the existence of the ground state is independent of the initial data, provided the seed theory is a CFT. The high-temperature density of states has Cardy-like behavior, rather than the Hagedorn growth characteristic of $T\bar{T}$-deformed theories. 
}

\maketitle

\section{Introduction}
The $T\bar{T}$ deformation \cite{Cavaglia:2016oda}, \cite{Smirnov:2016lqw} defines a flow on the space of two dimensional Euclidean quantum field theories. This flow, parametrized by $\mu$, defines a differential equation in terms of the partition function $Z$ that reads
\begin{equation}
\partial_{\mu} Z = \int \langle T\bar{T}(x)\rangle,
\end{equation}
where $\langle T\bar{T} \rangle$ is the expectation value of the so called $T\bar{T}$ operator. Written covariantly, it reads 
\begin{equation}
T\bar{T}(x) = T^{\mu\nu}T_{\mu\nu}-(T^{\mu}_{\mu})^{2}.
\end{equation}
If the theory at the origin of this flow, i.e. the undeformed theory is a conformal field theory, then there is a holographically dual description of this flow in terms of gravity in AdS$_{3}$ with a finite radial cutoff surface, on which Dirichlet boundary conditions are imposed. 

If we deform a CFT, the only energy scale in the problem is the one introduced by the $T\bar{T}$ deformation and so the $T\bar{T}$ flow equation is the Callan--Symanzik equation. 

An interesting fact about the holographically dual description to this flow is that it coincides with the radial development generated by the gravitational constraint equations in the bulk. In particular, in the limit of large central charge of the undeformed CFT, the flow equation for the partition function $Z[g]$ can be mapped on to the radial Hamiltonian constraint equation (for the radial wavefunction $\psi(g)$) in the bulk \cite{MMV}. Away from this limit, the expectation \cite{Mazenc:2019cfg}, \cite{Freidel:2008sh} is that the $T\bar{T}$ flow equation maps to the bulk radial Wheeler-de Witt equation (WdW), which is a quantization of the aforementioned constraint. The diffeomorphism constraint equations are simply a rewriting of the covariant conservation of the energy momentum tensor. In other words, $1/c$ is roughly playing the role of the Planck's constant in the bulk. This is the case of interest in the work to follow.

Despite the myriad lessons learned from the case where Dirichlet boundary conditions are imposed on this cutoff surface~\cite{MMV}, these boundary conditions are not elliptic. This lack of ellipticity prevents one from calculating quantities such as the propagator for metric fluctuations or the one loop determinant in the bulk theory.\footnote{The caveat being the situation where the extrinsic curvature is positive or negative definite. In that case, even with Dirichlet boundary conditions, the differential operator appearing in the kinetic term of the linearized theory is invertible~\cite{WittenBC}.} It turns out that conformal boundary conditions, where the trace of the extrinsic curvature of the cutoff surface and the conformal part of the metric adapted to it are fixed, are elliptic~\cite{2006math.....12647A} and would generically allow us to obtain the aforementioned quantities. We refer the readers to~\cite{WittenBC} for a more detailed explanation of this property. In this article, we will investigate the flow on the space of quantum field theories that AdS$_{3}$ gravity with these boundary conditions is dual to. 

Fixing Dirichlet boundary conditions in the bulk maps the on-shell action to the generating functional $\log{Z[g]}$, which can be exponentiated to obtain the partition function $Z[g]$. The functional dependence is on the boundary metric, which is the data that the Dirichlet boundary condition specifies at the cutoff surface. On the torus, this partition function turns out to depend only on the zero modes of the metric, namely the overall volume of the torus $V$ and the real and imaginary parts of the modular parameter $(\tau_{1},\tau_{2})$, i.e. 
\begin{equation}Z^{\mathbb{T}^{2}}[g]=Z(V,\tau_{1},\tau_{2}). 
\end{equation}

In order to change boundary conditions, one needs to perform a canonical transformation on the phase space of the bulk theory. This canonical transformation induces a certain Laplace transform of the $T\bar{T}$ deformed torus partition function, as a function of the volume:
\begin{equation}
\Gamma(T,\tau_{1},\tau_{2})= \int dV e^{-VT} Z(V,\tau_{1},\tau_{2}). 
\end{equation}
The resulting object is a quantum effective action $\log{\Gamma(T,\tau_{1},\tau_{2})}$ that depends on the trace mode of the energy momentum tensor, and behaves like the ordinary partition function in terms of its dependence on the conformal modes of the metric.
At the level of the free energy, it defines a Legendre transformation. The flow equation for $\Gamma(T,\tau_1,\tau_2)$ will be the quantity of interest in what follows, as will the thermodynamics of the new ensemble.

\subsection*{Organization of the Article}
This note is organized as follows. Section~\ref{sec:T2PartitionFunction} reviews the state of the science. In Section~\ref{sec:RewritingTheFlow}, we rewrite the $T\bar{T}$ flow equation above, changing variables from $\mu$ to a volume scale $V$. In these variables, the flow equation becomes the Wheeler-DeWitt equation of 3D gravity. To better demonstrate this correspondence from the gravity side, we study General Relativity in Constant Mean Curvature gauge. Then, in Section~\ref{sec:ImplementingConformalBCs}, we introduce the Legendre transform between $V$ and its canonically conjugate variable $T$, related to the trace of the extrinsic curvature that enters the boundary action. We show that the change in the boundary action due to the canonical transformation is exactly the right term to implement elliptic boundary conditions starting from a bulk with Dirichlet boundary conditions. We go on to derive and discuss the thermodynamics of this new ensemble in the low- and high-temperature regimes.

\section{The $\mathbb{T}^{2}$ partition function of $T\bar{T}$ deformed CFT$_{2}$}
\label{sec:T2PartitionFunction}
In this article, we will parameterize the family of $T\bar{T}$-deformed solutions by $\lambda$, the dimensionless $T\bar{T}$ coupling. The seed theory, i.e. the undeformed theory at $\lambda=0$, is a conformal field theory on the torus with central charge $c$. The modular parameter of the torus in question is denoted $\tau= \tau_1+i\tau_2$. The corresponding 2D line element reads:
\begin{equation}
ds^{2}= |dx+ \tau dy|^{2}.
\end{equation} 
where $x, y$ have period $2\pi R$. 
We would like to switch to units where we measure lengths in terms of the coupling constant of the $T\bar{T}$ deformation, $\mu$, that carries dimensions of length squared. 
Then, we can introduce the dimensionless version of this coupling: 
\begin{equation}
\lambda = \frac{\mu}{R^{2}},
\end{equation}
so that $R=\sqrt{\frac{\mu}{\lambda}}$. The volume of the torus is now given by:
\begin{equation}
V = \int d^{2}x \sqrt{g}= 4\pi^{2}\mu \frac{\tau_{2}}{\lambda}.
\end{equation}

In this language, the $T\bar{T}$ flow equation reads \cite{Aharony_2019}:
\begin{equation}
\partial_{\lambda}Z(\lambda,\tau_1,\tau_2)= \left(\frac{\tau_{2}}{4}\left(\partial^{2}_{\tau_{1}}+\partial^{2}_{\tau_{2}}\right)+ \frac{\lambda}{2}\partial_\lambda \partial_{\tau_2}-\frac{1}{2\tau_{2}}\lambda\partial_{\lambda}\right)Z(\lambda,\tau_1,\tau_2). \label{flow_eq}
\end{equation}
This equation is analogous to a diffusion equation, and the partition function $Z(\lambda,\tau_1,\tau_2)$ admits a representation in terms of an analogue of the heat kernel: 
\begin{equation}
Z(\lambda,\tau_1,\tau_2)= \frac{\tau_{2}}{\pi \lambda}\int_{\mathbb{H}} \frac{\textrm{d}^{2}\sigma}{\sigma^{2}_{2}} e^{-\frac{1}{\lambda \sigma_{2}}\vert \sigma - \tau\vert^{2}}Z_{CFT}(\sigma_{1},\sigma_{2}).
\end{equation}
Here, $Z_{CFT}(\sigma_{1},\sigma_{2})$ in the integrand is the initial condition $Z(\lambda=0,\tau_1,\tau_2)= Z_{CFT}(\tau_{1},\tau_{2})$, i.e. the flow originates from a CFT. It has been shown that the solutions to this differential equation are invariant under modular transformations, like the seed CFT \cite{Aharony_2019}, \cite{Datta:2018thy}. 
Also note that the domain of integration is the upper half plane $\mathbb{H}$. This expression appeared in \cite{Dubovsky:2018bmo}, \cite{Hashimoto:2019wct}, \cite{Callebaut:2019omt} and \cite{Mazenc:2019cfg}. Note that it can also be derived from the prescription of \cite{Tolley:2019nmm}, meaning that it can be seen as an expression for the path integral of 2d ghost-free massive gravity coupled to a conformal field theory. $Z_{CFT}$ can then be written as a partition sum
\begin{equation}
Z_{CFT}= \sum_{n}e^{-\tau_{2} E_{n}+i \tau_{1} J_n },
\end{equation}
where
\begin{equation}
E_{n}= \Delta_{n}+\bar{\Delta}_{n}-\frac{c}{12}, \qquad  J_n = \Delta_{n}- \bar{\Delta}_{n}.
\end{equation}
Here $\Delta_n$ and $\bar{\Delta}_n$ are the left and right conformal dimensions of the conformal field theory and $c$ is its central charge. We can find the $T\bar{T}$-deformed partition function by assuming that a similar form will hold:
\begin{equation}
Z(\lambda,\tau_1,\tau_2) = \sum_{n} e^{-\tau_{2}\mathcal{E}_{n}(\lambda)+i \tau_{1} J_{n} }.
\end{equation}
The reason why this form of the deformed partition function is justified (in particular, why the term involving $J_{n}$ remains unmodified) is tied to the fact that the $T\bar{T}$ deformation preserves translation invariance. Then, \eqref{flow_eq} leads to the following equation for the deformed energy levels $\mathcal{E}_{n}$:
\begin{equation}
2\lambda\mathcal{E}_{n} \partial_{\lambda}\mathcal{E}_{n}+ 4\partial_{\lambda}\mathcal{E}_{n}+\mathcal{E}^{2}_{n} = J_n^2. 
\end{equation}

This equation can be solved to obtain the deformed energy levels, given by:
\begin{equation}
\mathcal{E}_{n}(\lambda)= \frac{-2 + \sqrt{4+ 4\lambda E_n  + \lambda^2 J_n^2}}{\lambda}.
\end{equation}
The branch of the square root is selected so that $\mathcal{E}_n(\lambda \to 0) = E_n$. Our fellow $T\bar{T}$ aficionados should note that this expression lacks the traditional factors of $R$ as in Eq. (1.9) of~\cite{MMV}.
In order to reinstate the $R$ dependence, we take $\lambda \mapsto \frac{\mu}{R^{2}}$, so we find
\begin{equation}
\mathcal{E}_{n}(R)= \frac{2R^{2}}{\mu}\left(-1+\sqrt{1+\frac{\mu^2 J_n^{2}}{4R^{4}}+\frac{\mu E_{n}}{R^{2}}}\right)=R\tilde{E}_{n}.
\end{equation}
 In other words, $\mathcal{E}$ is the product of $R$ and what would normally be considered as the deformed energy levels. In the analysis which follows, we derive expressions in terms of $\mathcal{E}_n(R)$ rather than $\tilde{E}_n$, to keep expressions simple.

\section{Rewriting the flow equation}
\label{sec:RewritingTheFlow}
We can rewrite the flow equation in terms of the volume $V=4\pi^{2}\mu \frac{\tau_2}{\lambda}$.
\begin{equation}
\tau^{2}_{2}\left(\partial^{2}_{\tau_{1}}+\partial^{2}_{\tau_{2}}\right)Z(V,\tau_{1},\tau_{2})+V^{2}\left(\frac{1}{\pi^{2}\mu}\partial_{V}-\partial^{2}_{V}\right)Z(V,\tau_{1},\tau_{2})=0.
\end{equation}
This form of the equation will prove to be useful in making a connection to the Wheeler-de Witt equation of three dimensional quantum gravity. In particular, if we look at the object: 
\begin{equation}
\psi(V,\tau_{1},\tau_{2})= e^{-\frac{V}{2\pi^{2} \mu}}Z(V,\tau_{1},\tau_{2}),
\end{equation}
the equation it satisfies:

\begin{equation}
\tau^{2}_{2}(\partial^{2}_{\tau_{1}}+\partial^{2}_{\tau_{2}})\psi(V,\tau_{1},\tau_{2})+V^{2}\left(\frac{1}{4\pi^{4}\mu^{2}}-\partial^{2}_{V}\right)\psi(V,\tau_{1},\tau_{2})=0, \label{eqn:flow_v}
\end{equation}
 is identical to the Wheeler-de Witt equation in three dimensions with negative cosmological constant, in constant mean curvature gauge. We now define
 	\begin{align}
 		\Lambda \equiv \dfrac{1}{16\pi^4\mu^2}.
 	\end{align}
Henceforth, we work interchangeably with $\Lambda$ and $\mu$ in order to keep equations simple and draw important connections to the literature.

Note that this exercise is the finite-$c$ equivalent of deriving the trace flow equation. 
 \subsection{General Relativity in Constant Mean Curvature gauge}
The Arnowitt--Deser--Misner (ADM) Hamiltonian and momentum constraints are \cite{Arnowitt:1962hi}
 \begin{equation}
 \mathcal{H}=\frac{1}{\sqrt{g}} g_{i j} g_{k l}\left(\pi^{i k} \pi^{j l}-\pi^{i j} \pi^{k l}\right) - \sqrt{g}(R-2 \Lambda)
\end{equation}
\begin{equation}
\mathcal{H}_{i}=-2\nabla_{j}\pi^{j}_{i}=0.
\end{equation}

At the level of an action, we can implement these constraints with the term
\begin{equation}
H_{Tot} = \int\textrm{d}^{D}x\left( N(x) \mathcal{H}(x) + \xi^{i}(x) \mathcal{H}_{i}(x)\right)=H(N)+H_{i}(\xi^{i}).
\end{equation}
where the spacetime dimension of the bulk is $D+1$.

We would like to fix the mean curvature of the hypersurface to be constant. To do that we start by
splitting the conjugate momentum $\pi^{ij}$ into traceless and trace components, and define the metric $g_{ij}$ as a conformal rescaling of a constant-curvature counterpart $\bar{g}_{ij}$ via dilaton $\phi(x)$:
\begin{equation} \label{eqn:FieldDecomp}
\pi^{ij}= \sigma^{ij}+\frac{1}{D} \textrm{tr}\pi g^{ij},\,\,\qquad g_{ij}=e^{2\phi(x)}\bar{g}_{ij}.
\end{equation}
The gauge fixing condition imposes the constancy of $T$ defined as: 

\begin{equation}
T = \frac{2}{D}\frac{\textrm{tr}\pi}{\sqrt g}, \,\, \nabla_{i}T=0
\end{equation}
Specialising the the case of three dimensional gravity, in this gauge, the Hamiltonian constraint becomes:
\begin{equation}
\mathcal{H}_{CMC}=-\frac{1}{2} \sqrt{\bar{g}} e^{2 \phi}\left(T^{2}-4 \Lambda\right)+\sqrt{\bar{g}} e^{-2 \phi} \sigma^{i j} \sigma_{i j}+2 \sqrt{\bar{g}}\left[\bar{\Delta} \phi-\frac{1}{2} \bar{R}\right]=0,
\end{equation}
where ``barred'' quantities are defined in terms of $\bar{g}_{ij}$. Integrating, we find:
\begin{equation}
h_{CMC} = \int \textrm{d}^{D}x\,\,\mathcal{H}_{CMC}=-V^2(T^{2}-4\Lambda) + \tau^{2}_{2}\delta_{ab}p^{a}_{\tau}p^{b}_{\tau}=0.
\end{equation}
Note that classically, the condition for the existence of solutions is $T^{2}\geq 4\Lambda$. We will see how the quantum theory overcomes this bound. 

This Hamiltonian appeared first in \cite{doi:10.1063/1.528475}. For a review of classical and quantum gravity in 2+1 dimensions, see \cite{CarlipNotes}.

\subsection{The spectrum}
On the reduced phase space, the dynamics of the theory is finite dimensional. It can be quantized as such, and a quantum mechanical theory is obtained. The wavefunctions of interest depend on the modular parameters of the torus, as well as the volume. The corresponding momenta act as derivatives with respect to the conjugate variables: 
\begin{equation}
\hat{T}\psi(V,\tau_{1},\tau_{2})= -\partial_{V}\psi(V,\tau_{1},\tau_{2}),
\end{equation}
\begin{equation}
\hat{p}_{\tau_{a}}\psi(V,\tau_{1},\tau_{2}) = -\partial_{\tau_{a}}\psi(V,\tau_{1},\tau_{2}). 
\end{equation}

With these conventions, the global Hamiltonian constraint equation reads:
\begin{equation}
\hat{h}_{CMC}\,\psi(V,\tau_{1},\tau_{2})= \left[\tau^{2}_{2}(\partial^{2}_{\tau_{1}}+\partial^{2}_{\tau_{2}})-V^{2}\left(\partial^{2}_{V}-4\Lambda\right)\right]\psi(V,\tau_{1},\tau_{2})=0.
\end{equation}
Note that the ordering of the $V^{2}\hat{T}^{2}$ and $\tau^{2}_2\delta_{ab}\hat{p}^a \hat{p}^b$ terms are picked automatically by the rewriting of the $T\bar{T}$ flow equation \eqref{eqn:flow_v}. The relationship between radially quantized 3d quantum gravity and the $T\bar{T}$ deformation of CFTs was first noted in \cite{Freidel:2008sh}, and explained further in \cite{MMV}, \cite{Tolley:2019nmm}. Such a connection is anticipated in higher dimensional generalizations of the $T\bar{T}$ flow as well, see  \cite{Caputa:2019pam,Belin:2020oib,Hartman:2018tkw}.\footnote{Note that this quantized flow equation is not an approximation of a functional differential equation, i.e. a minisuperspace approximation, but rather an exact expression because of the particular choice of gauge.}

Also, this equation involves only partial derivatives with respect to global quantities $(V,\tau_1,\tau_2)$. In the bulk, this is a direct consequence of the CMC gauge condition. At large $c$, the sphere partition function computed in~\cite{Donnelly:2018bef},~\cite{Caputa:2019pam} also depends only on a global quantity, i.e. the radius of the sphere. At finite $c$, if we relate the radial wavefunction with the partition function, then the CMC gauge fixing also leads to an ODE involving derivatives with respect to the radius only \cite{Donnelly:2019pie}. 
On the field theory side, this can be seen as a consequence of the calculation in \cite{Dubovsky:2018bmo}, and it is unclear as to why these two facts lead the same phenomenon.

If we take the following ansatz for the wavefunction:
\begin{equation}
\psi(V,\tau_{1},\tau_{2})= e^{-2\sqrt{\Lambda}V}\sum_{n} e^{- \tau_{2}\mathcal{E}_{n}(\frac{\tau_{2}}{\sqrt{\Lambda}V})+i \tau_{1}J_{n}},
\end{equation}
then we can recover the partition sum discussed in the previous sections,
\begin{equation}
\mathcal{E}_{n}=\frac{V}{2\pi^2 \mu \tau_{2}}\left(-1+ \sqrt{1+\frac{4\pi^2 \mu \tau_{2}}{V}E_{n}+\frac{4\pi^{2}\mu^{2}\tau_2^{2}}{V^{2}}J_{n}^{2}}\right). 
\end{equation}
This is simply a rewriting of the expressions we had above, recall the identification:
\begin{equation}\lambda= 4\pi^{2} \mu \frac{\tau_{2}}{V}. \notag
\end{equation}
\subsection{Relationship to Jackiw--Teitelboim gravity}
In this section, we will briefly note what happens when one of the cycles of the torus degenerates. From the perspective of the wavefunction, this restriction is imposed as the condition:
\begin{equation}
\partial_{\tau_{1}}\psi_{JT}= 0,\, \,\qquad \psi_{JT}(V,\tau_2)=\psi(V,\tau_{1}=0,\tau_{2}).
\end{equation}
The constraint equation now reads:
\begin{equation}
\Big(\tau^{2}_{2}\partial^{2}_{\tau_{2}}-V^{2}(\partial^{2}_{V}-4\Lambda)\Big)\psi_{JT}(V,\tau_{2})=0.
\end{equation}
Just as the wavefunction in the three dimensional theory can be related to a partition function, the same is true in the dimensionally reduced case:
\begin{equation}
\psi_{JT}(V,\tau_{2})= e^{-2\sqrt{\Lambda} V} Z_{JT}(V,\tau_{2}).
\end{equation}
This object satisifes an equation identical to the one in appendix B of \cite{FiniteCutoffJT}: 
\begin{equation}
V^{2}\left(4\sqrt{\Lambda} \partial_{V}-\partial^{2}_{V}\right)Z_{JT}+\tau^{2}_{2}\partial^{2}_{\tau_{2}}Z_{JT}=0.
\end{equation}
Further, if we take the ansatz:
\begin{equation}
Z_{JT}(V,\tau_{2})=\sum_{n} g\left(\frac{\tau_{2}}{\sqrt{\Lambda}V}\right) e^{-\tau_{2} \mathcal{E}_{n}(\frac{\tau_{2}}{\sqrt{\Lambda}V})},\,\qquad g\left(\frac{\tau_{2}}{\sqrt{\Lambda}V}\right)=1,
\end{equation}
we find the energy levels obtained in (1.2) of \cite{FiniteCutoffJT}:
\begin{equation}
\mathcal{E}^{\pm}_{n}(V,\tau_{2})=\frac{V}{2\pi^2\mu \tau_2}\left(1 \mp \sqrt{1 -  \frac{\tau_{2}}{V} E_{n}}\right).
\end{equation}
If the spectrum of the undeformed theory is continuous, we can write the general solution as:
\begin{equation}
Z_{JT}(V,\tau_{2})=\int_{0}^{\infty} d E \,  \rho_{+}(E) \, e^{-\tau_{2} \mathcal{E}^{+}(V, \tau_{2})}+\int_{-\infty}^{\infty} d E \,\rho_{-}(E) \, e^{-\tau_{2} \mathcal{E}^{-}(V, \tau_{2})}.
\end{equation}
One can find a density of states which accommodates both branches of $\mathcal{E}$, i.e. we can choose 
\begin{equation}
\rho_+(E) = \sinh\left(2\pi\sqrt{2C E}\right) ,\qquad
\rho_{-}(E)=\left\{\begin{array}{cc}
-\sinh (2 \pi \sqrt{2 C E}), & 0<E<\frac{1}{8 \lambda} \\
\widehat{\rho}(E), & E<0
\end{array}\right.
\end{equation}
where $C$ is related to the boundary value of the dilaton, and $\hat{\rho}(E)$ is an arbitrary function. A more in-depth analysis of this solution is available in \cite{FiniteCutoffJT}.

\section{Implementing conformal boundary conditions}
\label{sec:ImplementingConformalBCs}
For a review on boundary conditions in Euclidean gravity, see~\cite{WittenBC}. We will be interested in polarized boundary conditions. These are imposed by specifying a Lagrangian submanifold in the theory's phase space. On such a submanifold, the symplectic form $\Omega$ vanishes, i.e.
\begin{equation}
\Omega = \int \textrm{d}^{D}x\left( \delta \pi^{ij}\wedge \delta g_{ij}\right),	\,\qquad\, \Omega\vert_{\mathcal{L}}=0.
\end{equation}
 
 The Lagrangian submanifold corresponding to the Dirichlet boundary conditions is specified as follows:

\begin{equation}
\mathcal{L}_{Dirichlet}=\left\{ (g_{ij};\pi^{ij}):\, \pi^{ij}(g)=\frac{\delta S[g]}{\delta g_{ij}}\right\}.
\end{equation}
In this picture, the conjugate momentum to the metric is the quasilocal stress-energy tensor of Brown and York.
In order to do perturbation theory, we will need to compute the propagator. This exercise involves inverting the second order differential operator appearing in the kinetic term of the action for fluctuations of the metric to quadratic order. Ellipticity is the requirement that the space of zero modes of this operator is at most finite dimensional, and it ensures (among other properties) that the leading-in-momentum component of the kinetic operator for metric fluctuations is invertible. Dirichlet boundary conditions generically run afowl of this requirement. Alternatively, we can use the so-called ``conformal'' boundary conditions \cite{2006math.....12647A}, where we specify the following Lagrangian submanifold: 
\begin{equation}
\mathcal{L}_{Conformal} = \bigg\{ (g_{ij};\pi^{ij})= \left(\bar{g}_{ij}(\tau_a), V; \sigma^{ij}(p^{a}),T\right): \,\sigma^{ij}(\tau_a)=\frac{\delta S}{\delta \bar{g}_{ij}}, V(T)= \frac{\delta S}{\delta T} \bigg\},
\end{equation}
where $S = S(T,\tau_1,\tau_2)$. This choice of boundary conditions is elliptic and thus has a well-defined propagator. In switching from Dirichlet to conformal boundary conditions, we note the following:
\begin{align}
\int\textrm{d}^{D}x \,\, \pi^{ij}\delta g_{ij} &= \int \textrm{d}^{D}x\left(\sigma^{ij}+\frac{1}{D}g^{ij}\textrm{tr}\pi \right)\delta g_{ij}.
\end{align}
The trace component in Constant Mean Curvature gauge simplifies to
\begin{equation}
\frac{1}{D}\int \textrm{d}^{D}x \left(\textrm{tr}\,\pi \, g^{ij}\delta g_{ij}\right)=\frac{2}{D}\int \textrm{d}^{D}x \left(\frac{\textrm{tr}\pi}{\sqrt{g}} \right)\delta \sqrt{g} \,\, \stackrel{CMC}{=} \,\, \int {\rm d}^Dx \,\, T \delta V,
\end{equation}
and the tracelessness of $\sigma^{ij}$ implies
\begin{align}
\int \textrm{d}^{D}x \,\, \sigma^{ij}\delta g_{ij}
&= \int \textrm{d}^{D}x \,( \sigma^{ij}\delta \bar{g}_{ij}),
\end{align}
where $\bar{g}_{ij}$ is as defined in~(\ref{eqn:FieldDecomp}). Then: 
\begin{align}
\Omega &= \int \textrm{d}^{D}x (\delta \bar{g}_{ij}\wedge \delta \sigma^{ij}) + \delta V\wedge \delta T = \delta \left(\int \textrm{d}^{D}x (\sigma^{ij}\delta \bar{g}_{ij})+T\delta V\right)    \\[0.5em]
&= \delta\left (\int \textrm{d}^{D}x (\sigma^{ij}\delta \bar{g}_{ij})+T\delta V-\delta (V T)\right)=\delta (p^{a}\delta \tau_{a}-V\delta T)
\end{align}
This is equivalent to a canonical transformation generated by the following shift in the action\footnote{Note that this canonical transformation is identical to the one used in the symmetry trading map of~\cite{Budd:2011er}.}:
\begin{equation}
G = VT. 
\end{equation}
More specifically, we are arguing that 
\begin{equation}
S_{GHY}-S_{Conf} = G,\qquad S_{Conf}=\frac{1}{D}S_{GHY}.
\end{equation}
We will now show this explicitly. In CMC gauge, the standard Gibbons-Hawking-York boundary term in the action takes the form:
\begin{equation*}
S_{GHY}= -2\int \textrm{d}^{D}x \sqrt{g}\, {\rm tr }K = -\frac{2}{1-D}\int \textrm{d}^{D}x\, {\rm tr }\,\pi.
\end{equation*}
If we now plug in the form of $G$ it becomes clear that
\begin{equation*}
S_{GHY}-G = -2\left(\frac{1}{1-D}+\frac{1}{D}\right)\int \textrm{d}^{D}x \, \textrm{tr}\,\pi=- \frac{2}{D(1-D)} \int \textrm{tr}\,\pi,
\end{equation*}
and since
\begin{equation}
- \frac{2}{D(1-D)} \int \textrm{tr}\,\pi = -\frac{2}{D}\int \textrm{tr}\,K = \frac{1}{D}S_{GHY},
\end{equation}
we thus conclude that the relevant boundary term for the new ensemble is exactly the one prescribed by conformal boundary conditions, i.e. $S_{Conf}$. Note that the change of boundary conditions involving a Legendre transformation is much in keeping with the lesson of \cite{Witten:2001ua}.
We should rewrite our ``wavefunction'' $\psi$ by making the Laplace transform explicit:
\begin{align}
\Psi(T,\tau_{1},\tau_{2})	&= \int \textrm{d}V \, e^{-VT} \psi(V,\tau_{1},\tau_{2}), \label{eqn:PsiLaplace} \\
\Gamma(T,\tau_{1},\tau_{2})	&= \int \textrm{d}V e^{-V T} Z(V,\tau_{1},\tau_{2}) \label{eqn:GammaLaplace}
\end{align}
where we define $\Gamma(T,\tau_1,\tau_2)$ to be the partition function for this new ensemble.\footnote{Note that this expression looks very similar to the one appearing in \cite{Cottrell:2018skz}, except we only integrate over the (zero mode of the) Weyl factor of the metric, and as such our Laplace transform is only partial. Our aims are also different from those of the authors of~\cite{Cottrell:2018skz}.} 
We can compute the correlation functions of the Weyl mode from taking successive $T$ derivatives of this object. In this way, it is similar to the dilaton effective action that features in \cite{Komargodski:2011vj}.   

The Legendre transform of the Hamiltonian constraint yields modified flow equations for $\Psi$ and $\Gamma$:
\begin{equation}\label{eqn:TFlowEqnPsi}
\hat{h}_{CMC}\,\Psi(T,\tau_{1},\tau_{2}) =\bigg( -2T\partial_{T}-2\partial_{T}T - (T^{2}-4\Lambda)\partial^{2}_{T}+\tau^{2}_{2}\left(\partial^2_{\tau_1}+\partial^2_{\tau_2}\right)\bigg)\Psi=0.
\end{equation}
Similarly, the partition function $\Gamma$ must satisfy
    \begin{align}\label{eqn:TFlowEqnGamma}
        \left(-(4\sqrt{\Lambda} + T)T\partial_T^2 - 4(2\sqrt{\Lambda}+T)\partial_T - 2  + \tau_2^2(\partial_{\tau_1}^2 + \partial_{\tau_2}^2)\right)\Gamma = 0.
    \end{align}
The two equations above are the key results presented in this article.

\subsubsection*{Connection to Schr\"odinger equation with York curvature time}
    
As should be expected in the context of quantization, there is an ordering ambiguity implicit in the definition of the constraint equation (\ref{eqn:TFlowEqnPsi}). We fix the ambiguity by integrating by parts within the Laplace transform. 
This is identical to the prescription of~\cite{Moncrief:1989dx,HosoyaNakaoMaassForms}. However, if one relaxes this ordering, it is possible to rearrange $V$ and $T$ to find an interesting variant:
\begin{equation}\label{eqn:TFlow}
\hat{h}_{CMC}\,\Psi(T,\tau_{1},\tau_{2}) = \Big(-T\partial_{T} - (T^{2}-4\Lambda)\partial^{2}_{T}+\tau^{2}_{2}(\partial^{2}_{\tau_{1}}+\partial^{2}_{\tau_{2}})\Big)\Psi=0.
\end{equation}
This version of the flow equation has appeared in the literature in a different form, see e.g. (3.4) in~\cite{CarlipNotes} as well as~\cite{PhysRevLett.26.1656}. In the reduced phase space quantization of 3D gravity in a torus universe, with $T$ parametrizing the York curvature time slice, the analog of the Schr\"odinger equation is:
\begin{equation}
\frac{\partial \Psi}{\partial T} = \frac{-1}{\sqrt{T^{2}-4\Lambda}} \sqrt{\tau^{2}_{2}(\partial^{2}_{\tau_{1}}+\partial^{2}_{\tau_{2}})}\,\,\Psi.
\end{equation}
We refer to this expression as the ``York-Schr\"odinger equation.'' Now, note that
	\begin{align}
		(\partial_{\tau_1}^2 + \partial_{\tau_2}^2) \tau_2 = 0 \cdot \tau_2.	\notag
	\end{align}
Since $\tau_2$ is an eigenfunction of this linear differential operator with eigenvalue $0$, it is an eigenfunction of the square root of that operator with eigenvalue $0$. We can therefore rewrite the York-Schr\"odinger equation as follows:
\begin{align}
	- \sqrt{T^2-4\Lambda} \, \partial_T \underbrace{\left(\sqrt{\partial_{\tau_1}^2 + \partial_{\tau_2}^2} \, \Psi \right)}_{-\frac{\sqrt{T^2-4\Lambda}}{\tau_2}\partial_T \Psi }= \tau_2 (\partial_{\tau_1}^2 + \partial_{\tau_2}^2) \Psi .
\end{align}
Rearranging, we find the same flow in~(\ref{eqn:TFlow}). 

\subsection{Kernel for $\Gamma$ and exact solutions for $\Psi$}
Writing out~(\ref{eqn:GammaLaplace}) with $V$ in terms of $\lambda$ allows us to do the Laplace transform explicitly:
    \begin{align}
        \Gamma = \int_{0}^\infty d\left(\dfrac{\tau_2}{\sqrt{\Lambda}\lambda}\right) e^{-\frac{\tau_2}{\sqrt{\Lambda}\lambda}T} \frac{\tau_2}{\pi\lambda} \int_{\mathbb{H}} \dfrac{d^2\sigma}{\sigma_2^2} e^{- \frac{1}{\lambda \sigma_2} |\sigma - \tau|^2} Z_{CFT}(\sigma_1,\sigma_2) 
    \end{align}
One can perform the $\lambda$ integral to find:
    \begin{align} \label{eqn:GammaKernel}
        \Gamma = \dfrac{\sqrt{\Lambda}}{\pi T^2} \int_{\mathbb{H}} \dfrac{d^2\sigma}{\sigma_2^2} \dfrac{Z_{CFT}(\sigma_1,\sigma_2)}{\left(1 - \frac{\sqrt{\Lambda}}{\tau_2\sigma_2T}|\sigma - \tau|^2\right)^2}. 
    \end{align}
This kernel can be shown to satisfy the flow in~(\ref{eqn:TFlowEqnGamma}). A condition on the integrand which arises as a requirement for the convergence of the $\lambda$ integral is
    \begin{align} \label{eqn:GammaKernelConstraint}
        T \left(\dfrac{T}{4\sqrt{\Lambda}}-1\right) \tau_2^2 < \sqrt{\Lambda}(\sigma_1 - \tau_1)^2 + \sqrt{\Lambda}\left(\sigma_2 - \left(1 - \frac{T}{2\sqrt{\Lambda}}\right)\tau_2\right)^2,
    \end{align}
which determines a circle or an annulus in the $(\sigma_1,\sigma_2)$ plane, depending on the magnitudes and signs of different parameters. To derive the above constraint, one must separately consider $\sqrt{\Lambda} \gtrless 0$. When that square root is negative, one should write the Laplace transform over negative values of $\lambda$, to keep $V$ positive. The overall sign of the argument to the exponential being integrated must be negative, which gives the stated result. 

This constraint makes the convergence properties of $\Gamma$ somewhat subtle, as we have to redefine the integration over $d^2\sigma$ within only a subregion of the upper half-plane. However, because the integrand itself solves the differential equation (\ref{eqn:TFlowEqnGamma}), we can restrict the domain of integration however we like and it will not affect the actual flow.  

The solutions to the $V,\tau_1,\tau_2$ flow equation for $\psi$ in~(\ref{eqn:flow_v}) have been studied extensively in the literature, see~\cite{HosoyaNakaoMaassForms} for a helpful review. Separation of variables yields a complete set of solutions in that case. We can find solutions to our $\Psi$ flow equation (\ref{eqn:TFlowEqnPsi}) by a similar method. Set $\Psi = \alpha(T)\beta(\tau_1)\gamma(\tau_2)$, and divide by $\Psi$ on both sides to find
	\begin{align}
		2 +4 T \dfrac{\alpha'(T)}{\alpha(T)} + (T^2-4\Lambda) \dfrac{\alpha''(T)}{\alpha(T)} = \tau_2^2 \left(\frac{\beta''(\tau_1)}{\beta(\tau_1)} + \frac{\gamma''(\tau_2)}{\gamma(\tau_2)}\right) \equiv -P
	\end{align}
Where $P$ is dimensionless and constant in all parameters. We can rearrange the expression on the RHS to see further that
	\begin{align}
		 \frac{\beta''(\tau_1)}{\beta(\tau_1)} = \dfrac{P}{\tau_2^2} - \frac{\gamma''(\tau_2)}{\gamma(\tau_2)} \equiv -(2\pi J)^2
	\end{align}
Where $J$ is also constant in all parameters. This gives us $3$ separate equations which we can solve for $\alpha$, $\beta$, and $\gamma$. In fact, since the zeroth-order Maass form determines the form of the $\tau_1$ and $\tau_2$ dependence in both the $T$ and $V$ flows, our $\beta$ and $\gamma$ will take the same form as in~\cite{HosoyaNakaoMaassForms}. In particular, modular invariance requires $J \in \mathbb{Z}$, and we obtain:
	\begin{align} \label{eqn:PsiExactSolns}
		\alpha(T) &= \alpha_{(1)} \,_2 F_1\left(\dfrac{1}{4}(3 -\nu),\dfrac{1}{4}(3 + \nu); \dfrac{1}{2}; \dfrac{T^2}{4\Lambda} \right)	 & (\nu = \sqrt{1 - 4P})\notag\\
				&\qquad+ \alpha_{(2)} \dfrac{T}{2\sqrt{\Lambda}} \,_2F_1\left(\dfrac{1}{4}(5 -\nu),\dfrac{1}{4}(5 + \nu); \dfrac{3}{2}; \dfrac{T^2}{4\Lambda} \right) 	\notag\\[0.5em]
		\beta(\tau_1) &= \beta_{(1)} \, e^{2\pi i J\tau_1} + \beta_{(2)} \,  e^{- 2 \pi i  J \tau_1}	\\[0.5em]
		\gamma(\tau_2) &= \gamma_{(1)} \sqrt{\tau_2}\,  K_{\frac{1}{2}\nu}( 2\pi |J| \tau_2) 	\notag
	\end{align}
In the absence of a specific surface on which the CFT ``lives,'' it is difficult to impose any further constraints on the form of this solution. Nevertheless we can comment on its various properties, for example convergence. The form of $\alpha(T)$ is simply the appropriate solution to the hypergeometric differential equation. The defining series for $\,_2F_1$ around $T^2/4\Lambda=0$ happens to converge only for $T^2 < 4\Lambda$, with divergences at equality, but the region $T^2 > 4\Lambda$ is nevertheless accessible. This is because $\,_2 F_1(a,b;c;z)$ can be written as a linear combination of $\,_2 F_1$'s with the argument $z$ replaced with one of the other 5 cross-ratios involving $z$ and 1. In addition, $\gamma(\tau_2)$ is well-behaved for all $P$ even though the order becomes pure imaginary for $P > \frac{1}{4}$; this follows from standard properties of the modified Bessel functions. 

Regarding $\alpha$, one should note the difference with the analogous solution to the York-Schr\"odinger equation:
	\begin{align}
		\alpha_{YS}(T) &= \alpha_{(1)} \cos\left(\sqrt{P} \, {\rm arctanh}\left(\dfrac{T}{\sqrt{T^2-4\Lambda}}\right)\right)	+ \alpha_{(2)} \sin\left(\sqrt{P} \, {\rm arctanh}\left(\dfrac{T}{\sqrt{T^2-4\Lambda}}\right)\right)
	\end{align}
The behavior of $\beta$ and $\gamma$ is again unaffected. In addition, the apparent singular behavior in $\alpha(T)$ at $T^2 = 4\Lambda$ is a removable discontinuity.

\subsection{Ground state existence and asymptotic density of states}
\label{sec:GSExist}

We ultimately wish to study the thermodynamics of this new system, specifically the energy levels. However, a na\"ive approach will not work, because the energies in the $Z(V,\tau_1,\tau_2)$ ensemble depend nontrivially on $\tau_2$ and $V$. The flow equation for $\Gamma$ can be solved by separation of variables just as $\Psi$ can. In analogy with the expressions in~(\ref{eqn:PsiExactSolns}), we find the rather unenlightening solution:
    \begin{align}
        \alpha(T) = \alpha_{(1)} \,_2 F_1\bigg(\dfrac{1}{2}(3& -\nu),\dfrac{1}{2}(3 + \nu); 2; -\dfrac{T}{4\sqrt{\Lambda}} \bigg)   \notag \\[0.5em]
            & + \alpha_{(2)} G^{2\,0}_{2\,0}\left(-\dfrac{T}{4\sqrt{\Lambda}}\,\Bigg|\, \begin{matrix}
                -\frac{1}{2}(1 + \nu),\, -\frac{1}{2}(1 - \nu) \\
                -1,\, 0 
            \end{matrix}\right),
    \end{align}
where $\nu$, $\beta(\tau_1)$, and $\gamma(\tau_2)$ are the same as for $\Psi$. To gain some insight into the spectrum, we might instead consider the discrete version of the Laplace transform: 
    \begin{align}
        \Gamma(T,\tau_1,\tau_2) =  \sum_{n} e^{-n V T} Z(n V,\tau_1,\tau_2).
    \end{align}
Since the left-hand side is independent of $V$, the Hamilton-Jacobi equations tell us that 
    \begin{align}
        \partial_V \Gamma = 0 &= \sum_{n} n e^{-n V T} \Big( - T + \partial_{nV}   \notag Z(nV,\tau_1,\tau_2)\Big),  \\
    &\implies T = \dfrac{\sum_n n e^{-nVT} \partial_{nV} Z(nV,\tau_1,\tau_2)}{\sum_n n e^{-nVT}}.
    \end{align}
However, inverting the sums to study $V$ is not a tractable approach. Taking a direct ansatz for the energy levels fails similarly, as it requires solving nonlinear second-order differential equations. We are thus restricted to studying the thermodynamics of this ensemble in various limits. 
    
For $\tau_2 \gg 1$ and $\tau_1=0$, which can be viewed as the low-temperature limit, the Legendre transform can be performed explicitly. The inverse temperature is given by $\beta = \frac{\tau_2}{R}$ where $R$ is the radius of the torus. We will nevertheless proceed with our present conventions and treat $\tau_2$ as the inverse temperature. We will follow a line of reasoning similar to that in \cite{Datta:2018thy} which was aimed at extracting the density of states from $Z(\lambda, \tau_1,\tau_2)$. The leading term in the low temperature expansion of the partition function is:
\begin{equation}
Z(\tau_2,V)\vert _{\tau_{2}\gg 1}\sim \, e^{\frac{V}{2\pi^2\mu \tau_2}\left(1 \mp \sqrt{1 +  \frac{c\tau_{2}}{12 V}}\right).}.
\end{equation}
We now apply the Hamilton-Jacobi equation, i.e.
	\begin{align}
		0 &\equiv \dfrac{\partial \log\Gamma}{\partial V} = \dfrac{\partial \log Z}{\partial V} - T, 
	\end{align}
to find $V(T,\tau_2)$:

\begin{equation}
	V^{\pm}(T,\tau_2) = \frac{c\tau_2 }{24\sqrt{\Lambda}}\left( 1 \pm \sqrt{ 1 + \dfrac{4\Lambda}{T(T-4\sqrt{\Lambda})}} \right),
\end{equation}
which is real provided $\frac{T}{\sqrt{\Lambda}}<0$ or $\frac{T}{\sqrt{\Lambda}}> 4$. Then, by plugging this back into the low temperature limit, we obtain the asymptotic form of $\Gamma$:
\begin{equation}
\Gamma(T,\tau_2)\vert_{\tau_2\gg 1}= e^{-\tau_{2}\mathfrak{E}_o},
\end{equation}
where 
\begin{equation}
\mathfrak{E}_o^{\pm}(\mathcal{X}) = -\frac{c}{12}\left(1-\frac{1}{2} \mathcal{X}\right) \left( 1 \pm\sqrt{ 1 + \dfrac{4}{\mathcal{X}(\mathcal{X}-4)}} \right) - \sqrt{\dfrac{4}{\mathcal{X}(\mathcal{X}-4)}}.
\end{equation}
Here, we write the effective ground state energy in terms of the dimensionless quantity
$$\mathcal{X} = \frac{T}{\sqrt{\Lambda}}.$$

We note that this quantity is real whenever $V(T,\tau_2)$ is real. Moreover, it does not depend on $\tau_2$, unlike $\mathcal{E}_o$ and $V$. This means that $\mathfrak{E}_o$ can be properly interpreted as an energy. Note also that $\Gamma$ is invariant under modular transformations. To understand this, we restate~(\ref{eqn:GammaLaplace}),
\begin{equation} 
\Gamma(T,\tau_1,\tau_2)= \int dV\, e^{-VT}Z(V,\tau_1,\tau_2), \notag
\end{equation}
and note that $T,$ $V$, and $Z(V,\tau_1,\tau_2)$ are all modular invariant.\footnote{One might reasonably be concerned that the constraint in~(\ref{eqn:GammaKernelConstraint}) ruins this argument, because $\Gamma$ can no longer be written in terms of $Z(V,\tau_1,\tau_2)$. In this article, we consider~(\ref{eqn:GammaLaplace}) to be the definition of $\Gamma$. Whether it converges only under a reduced class of conditions is a question we relegate to future work.}
Thus we can apply an S transformation that takes $\tau_2\rightarrow \frac{1}{\tau_2}$ and obtain the high temperature limit of $\Gamma(T,\tau_2)$:
\begin{equation}
\Gamma\left(T,\frac{1}{\tau_{2}}\right)\bigg\vert _{\tau_2\gg 1}= e^{\frac{\mathfrak{E}_o}{\tau_2}}.
\end{equation}
We can take the inverse Laplace transform of this quantity to obtain the density of states:
\begin{equation}
\rho(\epsilon)=\int^{i\infty} _{-i\infty} d\tau_2\, e^{\tau_2 \epsilon+\frac{\mathfrak{E}_o}{\tau_2}}.
\end{equation}
Using the saddle point approximation, we see Cardy-like growth at high energies: 
\begin{equation}
\rho(\epsilon)\sim e^{2\sqrt{\mathfrak{E}_o \epsilon}}.
\end{equation}
This is the regime that the high temperature expansion can accurately shed light on. Note that our findings are quite different from the case with Dirichlet boundary conditions, where one finds Hagedorn behaviour. Also note that $\mathfrak{E}_o \sim c$, so the above result also reflects the behaviour of the density of states in that limit. This is much akin to what happens in CFTs.

\section{Discussion}

In this note, we have derived a modification to the flow equation of the $T\bar{T}$ deformation which implements conformal boundary conditions in the bulk dual, rather than the Dirichlet boundary conditions of lore. By rewriting the flow in terms of a characteristic volume scale $V$, we identified it with the Wheeler-DeWitt equation of AdS$_3$ gravity. Then, starting from the bulk gravitational action with Dirichlet boundary conditions (i.e. with the Gibbons-Hawking-York boundary term), we change the variational problem to impose conformal boundary conditions. We show that in Constant Mean Curvature gauge, the term needed to shift between these pictures can be interpreted as a Legendre transform of the torus path integral. The resulting system has a ground state whose existence is independent of the CFT data once the deformation parameters are fixed. 

One may reasonably ask why we were interested in this problem in the first place, given that 3D gravity has no gravitons. In fact, if it can be well-defined, the graviton propagator will still contribute to any perturbative expansion in $G_N$, but will only enter as e.g. an internal line in a Feynman diagram. Holographically, the avatar of such an expansion is large-$c$ perturbation theory, which was the regime of interest here and in~\cite{MMV}. 
We wonder whether the lack of a $\mu$-dependent breakdown of the ground state of the Legendre transformed theory is due to the elliptic nature of the boundary conditions. 

We see many paths forward left to explore. Exact results for the full spectrum and density of states may be within reach, despite the troubling nonlinearities which generically arise in deriving $\mathfrak{E}_n(\mathcal{X})$ directly. In this vein, one option may be to integrate the kernel in (\ref{eqn:GammaKernel}) exactly, perhaps by taking advantage of its similarity to the structure of the Feynman propagator. Generalizations of this flow are also immediately available for the many generalizations of $T\bar{T}$. We are particularly interested in applying our procedure to dS holography using the flow prescribed in~\cite{dSdSTTbar}.

It would also be valuable to construct a string-theoretic analog of the flow in $T$, in the spirit of~\cite{Kutasov1,Kutasov2}. In those works, a deformation of the worldsheet sigma model implements a single-trace variant of the $T\bar{T}$ deformation in the putative 2D boundary CFT of the AdS$_3$-like spacetime. For the ``holographic'' sign of the deformation, their procedure yields spacetimes with naked singularities. We are curious whether a similar procedure is possible for our proposed flow, and if so, what the salient features are of the corresponding geometries. 

There is one overarching question which we hope to address in future work: where does the analogy with the BTZ phase space enter in this new picture? The answer is unclear, given the results presented. However, there are some rough hints of a correspondence worth mentioning at the level of discussion. 
In our case, we started by deforming a unitary CFT without a gravitational anomaly and thus $E_o = -\frac{c}{12}$ and $J_o=0$. However, if we start with a theory with a gravitational anomaly, and further assume that our expressions remain valid in this setting, then the expression one obtains for the volume is:
\begin{equation}
	V^{\pm}(T,\tau_2) = -\frac{E_o\tau_2 }{2\sqrt{\Lambda}}\left( 1 \pm \sqrt{1 - \left(\frac{J_o}{E_o}\right)^2}\sqrt{ 1 -  \dfrac{4\Lambda}{T(T-4\sqrt{\Lambda})}} \right).
\end{equation}
In order for this quantity to be real, one must separately consider $|\frac{J_o}{E_o}| \gtrless 1$, and then identify the range of $\frac{T}{\sqrt{\Lambda}}$ which keeps the argument to the square root nonnegative. This is eerily reminiscent of the bounds on angular momentum which arise in the BTZ phase space, but we will curtail our speculative commentary here. 

\section*{Acknowledgments}
We thank Eva Silverstein, Shouvik Datta, Gonzalo Torroba, Jorrit Kruthoff, Ronak Soni, G. Joaquin Turiaci, Zhenbin Yang, and Victor Gorbenko for their insightful comments.  The work of EAC is supported by the US NSF Graduate Research Fellowship under Grant DGE-1656518.
During early stages of this work, VS was supported in part by the Perimeter Institute for theoretical Physics. Research at Perimeter Institute is supported in part by the Government of Canada through the Department of Innovation, Science and Industry Canada and by the Province of Ontario through the Ministry of Colleges and Universities.


\bibliographystyle{utphys}
\bibliography{refs}

\end{document}